\documentclass[prl,final,floatfix,twocolumn,showpacs]{revtex4-1}

\usepackage{graphicx}
\usepackage{dcolumn}
\usepackage{bm}
\usepackage{amssymb}
\usepackage{natbib}
\usepackage{amsmath}
\usepackage{color}
\usepackage{datetime}
\usepackage{footnote}
\usepackage{bbold}
\usepackage{hyperref}
\usepackage[normalem]{ulem}
\usepackage{dcolumn}

\newcommand{\nogo}[1]{}
\newcommand{\suppl}[1]{28}
\newcommand{\method}{DCRT}

\newif\iffigs
\figstrue

\begin{document}

\title{Experimental characterization of quantum dynamics through
  many-body interactions}

\author{Daniel Nigg$^1$}
\author{Julio T. Barreiro$^{1}$}
\altaffiliation{Current address: Quantum Optics Group, Ludwig-Maximillians-University}
\email{Julio.Barreiro@uibk.ac.at}
\author{Philipp Schindler$^1$}
\author{Masoud Mohseni$^3$}
\email{mohseni@mit.edu}
\author{Thomas Monz$^1$}
\author{Michael Chwalla$^{1,2}$}
\author{Markus Hennrich$^{1}$}
\author{Rainer Blatt$^{1,2}$}

\affiliation{$^1$Institut f\"ur Experimentalphysik, Universit\"at
Innsbruck, Technikerstrasse 25, A--6020 Innsbruck,
Austria\\$^2$Institut f\"ur Quantenoptik und Quanteninformation
der \"Osterreichischen Akademie der Wissenschaften,
Technikerstrasse 21a, A--6020 Innsbruck, Austria\\
$^3$Research Laboratory of Electronics, Massachusetts Institute of Technology, Massachusetts 02139, USA}

\date{\today}

\begin{abstract}       
We report on the implementation of a quantum process tomography (QPT)
technique known as direct characterization of quantum dynamics (DCQD) applied 
on coherent and incoherent single-qubit processes in a system of trapped
$^{40}$Ca$^{+}$ ions. Using quantum correlations with an ancilla qubit, 
DCQD reduces exponentially the number of experimental configurations required 
for standard QPT. With this technique, the system's relaxation times
$\mathrm{T}_{1}$ and $\mathrm{T}_{2}$ were measured with a single experimental
configuration. We further show the first complete characterization of single-qubit 
processes using a single generalized measurement realized through multi-body 
correlations with three ancilla qubits. 
    
\end{abstract}


\maketitle

Characterization of quantum dynamics is an important primitive in
quantum physics, chemistry, and quantum information science for
determining unknown environmental interactions, estimating Hamiltonian
parameters, and verifying the performance of engineered quantum
devices. This has led to a major effort in developing tools for the
full characterization of quantum processes, known as quantum process
tomography (QPT). The standard approach for QPT is resource intensive, requiring
12$^{N}$ experimental configurations for a system of $N$
qubits~\cite{Chuang,SQPT_exp}, where each experimental configuration consists of the
preparation of input probe states and the measurement of process outputs. 
Using ancilla qubits but only joint separable measurements, the number of experimental configurations is
still 12$^{N}$~\cite{AAPT_Presti,AAPT,Mosheni_Resource}.
However, the use of many-body interactions to ancilla qubits
in the preparation and/or measurements can significantly decrease
this number to anywhere from 4$^{N}$ to a single
configuration depending on the nature and complexity of quantum
correlations~\cite{Mosheni_Resource}. Using two-body correlations DCQD 
requires up to 4$^{N}$ experimental configurations for full
quantum process tomography, and in particular only one experimental
setting for estimating certain parameters (e.g. relaxation times)~\cite{Mosheni_DCQD_2006, Mosheni_DCQD_2009}
(see Ref.~\cite{dcqd_exp,dcqd_single_photon} for partial and non-scalable implementations of DCQD).

Alternative tomographic methods such as randomized benchmarking ~\cite{bench},
selective and efficient QPT~\cite{selective_Paz, Paz} and compressed sensing for quantum process
tomography~\cite{shabani_1,shabani_2,compressed} have recently been
developed to overcome the exponential increase of the required
experimental configurations. Generally, these methods are tailored to estimate a polynomial number of
effective parameters, such as gate fidelity ~\cite{bench} or when we
can make a sparse quantum process/Hamiltonian assumption from a priori
knowledge about the quantum system~\cite{compressed}. For example, the estimation 
of the dynamical parameters $\mathrm{T}_{1}$ and $\mathrm{T}_{2}$
(longitudinal and transverse relaxation times~\cite{Chuang}) is a
task involving two non-commuting observables (e.g. $\sigma_{x}$
and $\sigma_{z}$) that cannot be measured simultaneously. These
parameters describe the influence of noise on atomic-, molecular- and spin-based
systems induced by the interaction with the environment. An
alternative approach based on DCQD, henceforth called Direct
Characterization of Relaxation Times (\method{}), enables the
measurement of both $\mathrm{T}_{1}$ and $\mathrm{T}_{2}$
\textit{simultaneously} with a single experimental
configuration~\cite{Mohseni_parameters}. 

In this work, we apply the DCQD technique and extensions on a system of trapped
$^{40}$Ca$^{+}$ ions. Single-qubit processes are reconstructed with four
experimental configurations using DCQD, and alternatively with just a
single configuration using a generalized measurement (GM). In
addition, we quantify the relaxation times T$_1$ and T$_2$ in our system with a
single configuration. This technique can also characterize more realistic
environments affecting not only the probe but also the ancilla qubit
collectively.\\

\begin{figure}[t]
\includegraphics{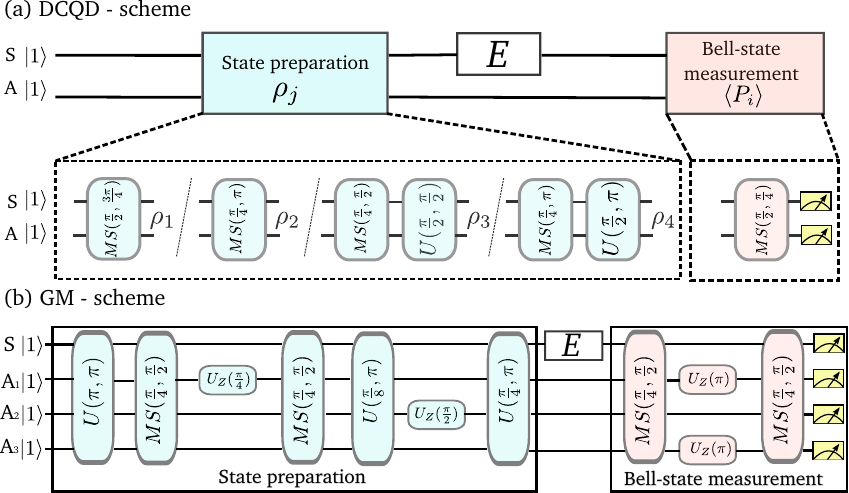}
\caption{\label{process_scheme}(color online) Procedure to characterize a single-qubit process
  with DCQD and a GM. In DCQD (a) each experimental
  configuration consists of the preparation of one of four input states $\rho_{j}$
  entangled between the system ion S and the ancilla ion A. The
  process $E$ is applied on S followed by a BSM on the output state
  $E(\rho_{j})$, which consists of a single MS operation followed by a projection 
  onto the computational basis. (b) Generalized measurement via many body
  interactions (see text).}
\end{figure}

Every process acting on a quantum system can be described by a
complete positive map $E$ mapping the input state $\rho$ onto the
output state $\rho'$. For a single qubit this can be written as
\begin{equation}
E: \rho \rightarrow \rho' = \sum\limits_{m,n=1}^{4}
\chi_{m,n}\ \sigma_{m}~\rho~\sigma_{n}^{\dagger},
\end{equation}
with $\sigma_{m}$, $\sigma_{n}$ the Pauli operators
$\{\mathbb{1},\sigma_{x},\sigma_{y},\sigma_{z}\}$ and $\chi$ a
semi-positive matrix containing complete information about the
process. In standard quantum process tomography (SQPT) the
process is applied to four input states and followed by full state
tomography of each output state, which for a trace preserving map consists of three
measurements, resulting in $4\times3=12$ experimental configurations. 
In DCQD these four input states are replaced by four
entangled states between the system qubit S and an ancilla
qubit A, and the state tomography is replaced by a single Bell-state
measurement (BSM), as shown in Fig.~\ref{process_scheme}(a), with a total of
$4\times1=4$ experimental configurations (Bell states $|\Psi^{\pm}\rangle$ and $|\Phi^{\pm}\rangle$ as defined in Table~\ref{table_states}).
The probabilities $p_{i,j}$ of measuring the
Bell-state projector $P_{i}$ for each input state $\rho_{j}$ shown in Table~\ref{table_states} are
determined, according to Refs.~\cite{Mosheni_DCQD_2006, Julio_DCQD_2010}, by
\begin{eqnarray}
& p_{i,j}& = \mathrm{Tr}(P_{i}\ E(\rho_{j}))  = \sum\limits_{m,n=1}^{4}\chi_{m,n}\ \Lambda_{m,n}^{i,j}\ \\  
&\Lambda_{m,n}^{i,j}& = \mathrm{Tr}(P_{i}(\sigma_{m}\otimes \mathbb{1})\rho_{j}(\sigma_{n}\otimes \mathbb{1})^{\dagger}). \nonumber
\end{eqnarray}
Therefore the process matrix $\chi$ can be calculated directly by
linear inversion of the matrix $\Lambda$. The set of input states
$\rho_{j}$ and the Bell-state projectors $P_{i}$ have to be determined
such that the 16 equations in Eq.~(2) are linearly independent, which
ensures that $\Lambda$ is invertible (Table~\ref{table_states}).

Our experiments were realized on a system consisting of
$^{40}$Ca$^{+}$ ions confined to a string in a linear Paul
trap~\cite{exp}. Each ion represents a logical qubit which is encoded
in the electronic levels $D_{5/2}(m = -1/2)=|0\rangle$ and $S_{1/2}(m
= -1/2)=|1\rangle$. Each experimental cycle consists of an
initialization of the ions in their internal electronic and motional
ground states followed by a coherent manipulation of the qubits and
finally a detection of the quantum state. State initialization is
realized by optical pumping into the $S_{1/2}(m = -1/2)$ state
after cooling the axial centre-of-mass mode to the motional
ground state. The manipulation of the qubits is implemented by
coherently exciting the $S_{1/2} \leftrightarrow D_{5/2}$ quadrupole
transition with laser pulses. Finally, the population of the qubit
states is measured by exciting the $S_{1/2} \leftrightarrow
P_{1/2}$ transition and detecting the fluorescence light, using
electron shelving~\cite{Dehmelt}. Our setup is capable of realizing
collective qubit rotations $U(\theta,\phi) =
\exp(-i\frac{\theta}{2}\sum_{i}[\sin(\phi)\sigma_{y}^{(i)}+\cos(\phi)\sigma_{x}^{(i)}])$
via a laser beam addressing the entire register as well as
M$\o$lmer-S$\o$renson entangling gates $MS(\theta,\phi) =
\exp(-i\frac{\theta}{4}[\sum_{i}\sin(\phi)\sigma_{y}^{(i)}+\cos(\phi)\sigma_{x}^{(i)}]^{2})$~\cite{ms,ms_exp}. 
Additionally we are able to perform single-qubit rotations on
the \textit{i}-th ion of the form $U_{Z}^{(i)}(\theta) =
\exp(-i\frac{\theta}{2}\sigma_{z}^{(i)})$ by an off-resonant laser
beam, which addresses individual ions. The input states for DCQD of
Table I are prepared by applying collective entangling operations and
qubit rotations as shown in Fig.~\ref{process_scheme}(a). For example,
the input state $\rho_{2}$ is created by the non-maximally entangling
operation $MS(\frac{\pi}{4},\pi)$. Our two-qubit entangling operation
generates Bell states with a fidelity of $\approx 99\%$ in 120~$\mu$s.

\begin{table}[t]
\begin{center}
\begin{tabular}{l l  l l}
\hline \hline
&  Input states $\rho_{j} = |\psi_{j}\rangle\langle\psi_{j}|$  & Bell-state basis \\
\hline
 &$|\psi_{1}\rangle = |00\rangle + |11\rangle$   &$|\Phi^{+}\rangle = |00\rangle + |11\rangle$ \\
 &$|\psi_{2}\rangle = \alpha|00\rangle + \beta|11\rangle$   &$|\Psi^{+}\rangle = |01\rangle + i|10\rangle$ \\
 &$|\psi_{3}\rangle = \alpha|++\rangle_{x} - \beta|--\rangle_{x}$   &$|\Psi^{-}\rangle = |01\rangle - i|10\rangle$ \\
 &$|\psi_{4}\rangle = \alpha|++\rangle_{y} - \beta|--\rangle_{y}$   &$|\Phi^{-}\rangle = |00\rangle - |11\rangle $\\
\hline
\hline
\end{tabular}
\caption{Input states and BSM basis used for the implementation of DCQD
  ($|\pm\rangle_{x} = \frac{|0\rangle \pm |1\rangle}{\sqrt{2}}$,
    $|\pm\rangle_{y} = \frac{|0\rangle \pm i|1\rangle}{\sqrt{2}}$). The determinant of 
$\Lambda$ in Eq.~(2) is maximized for $\alpha=\cos(\frac{3\pi}{8}$ 
 and $\beta=\exp(i\frac{\pi}{2})\sin(\frac{3\pi}{8})$) to ensure the invertibility~\cite{Julio_DCQD_2010}.
 The BSM is realized by a measurement with the projectors $P_{i}
  = \{|\Phi^{\pm}\rangle\langle\Phi^{\pm}|, |\Psi^{\pm}\rangle\langle\Psi^{\pm}|\}$.}
\label{table_states}
\end{center}
\end{table}

The BSM is experimentally realized by a maximally entangling operation
$MS(\frac{\pi}{2},\frac{\pi}{4})$, which maps from the Bell-state basis to the
computational basis $\{|00\rangle, |01\rangle, |10\rangle,
|11\rangle\}$, followed by individual-ion-resolving fluorescence
detection with a CCD camera. For example, consider the first input
state $\rho_{1} = |\Phi^{+}\rangle\langle\Phi^{+}|$. If the process
$E$ is the identity $\mathbb{1}$, the expectation value of the
BSM-projector $P_{1}$ is 1 which is equivalent to detecting both
ions in the state $|11\rangle$ after the BSM. If a bit flip occurs on
the system ion, the output state is then mapped onto the state
$|01\rangle$ by the BSM ($\langle P_{2}\rangle = 1$). The
considerations are similar for a phase-flip, or bit- and phase-flip
processes. Therefore, the diagonal elements $\chi_{m,m}$ of the
superoperator $\chi$ corresponding to $\mathbb{1}$, $\sigma_x$,
$\sigma_y$, and $\sigma_z$ are detected by a single input state in
combination with one BSM.

\begin{figure}[t]
\iffigs
\includegraphics[height=2.8cm,width=2.8cm]{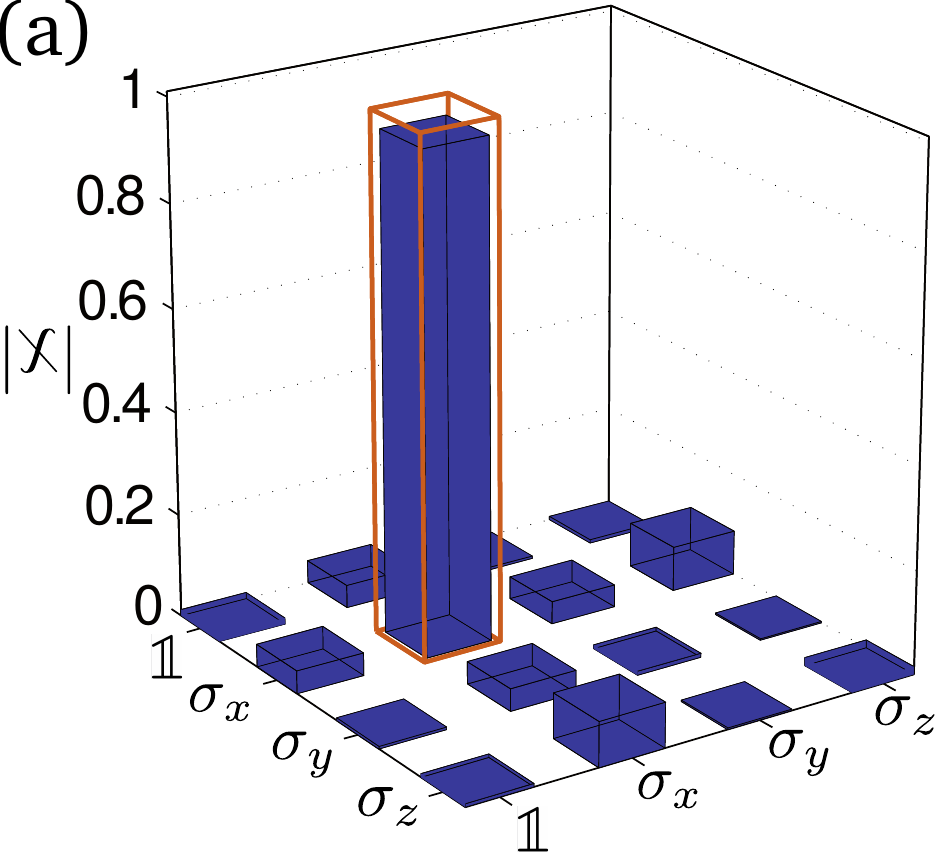}
\includegraphics[height=2.8cm,width=2.8cm]{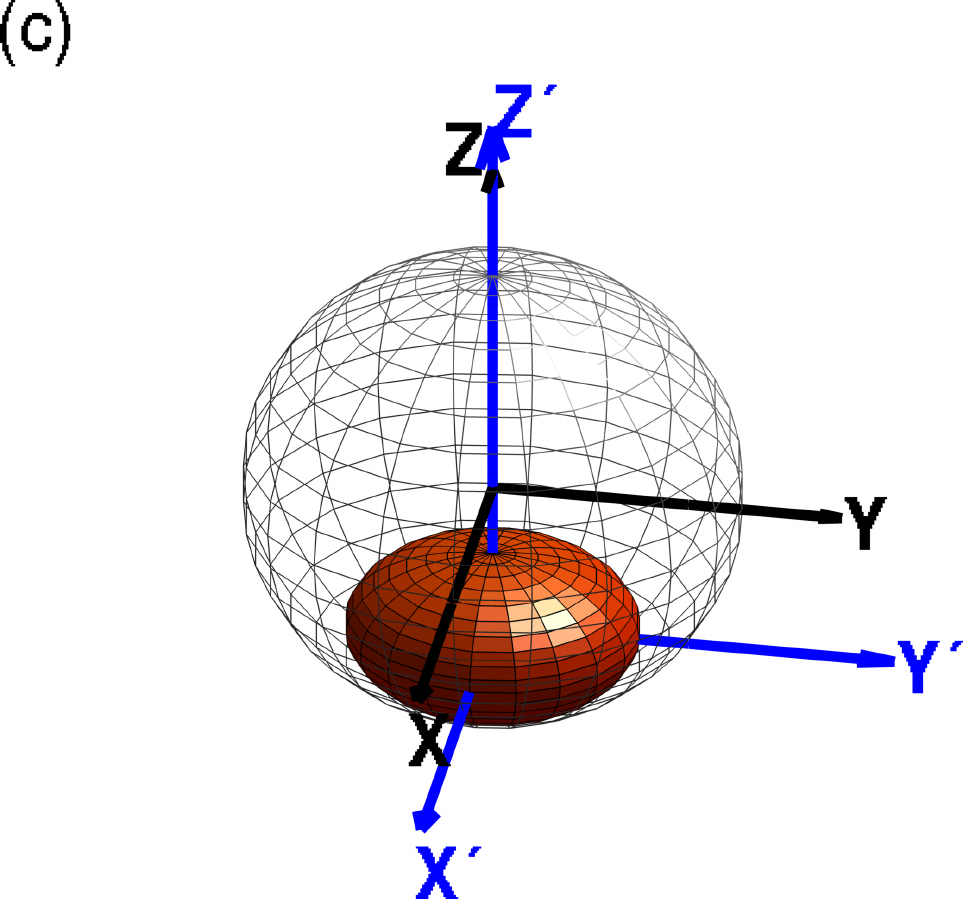}
\includegraphics[height=2.8cm,width=2.8cm]{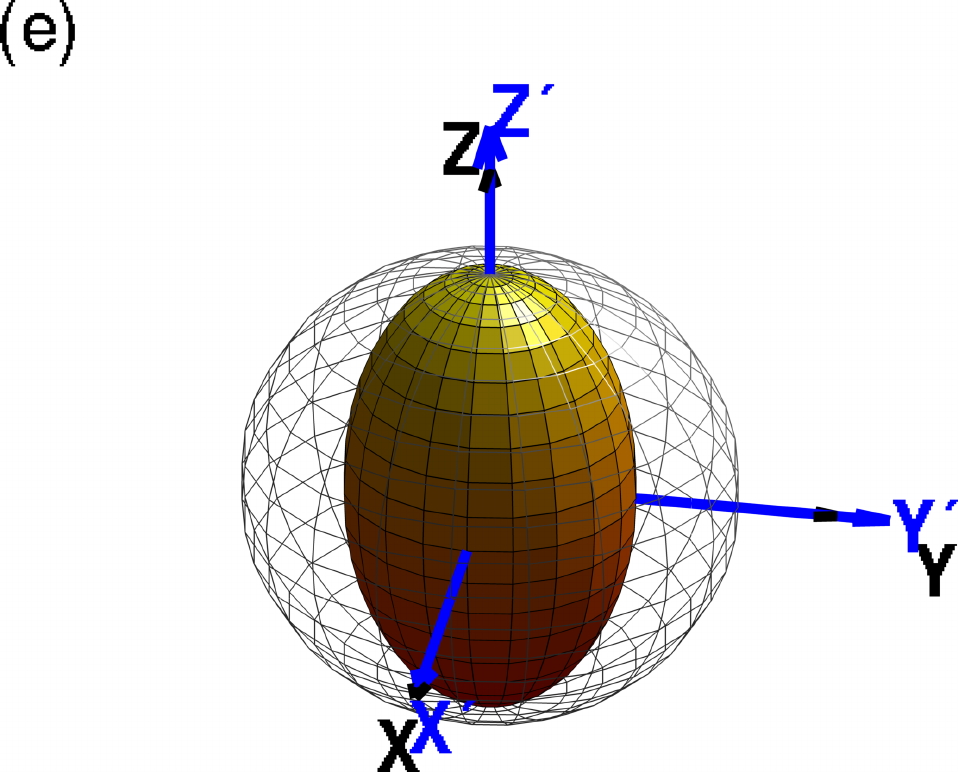}
\includegraphics[height=2.8cm,width=2.8cm]{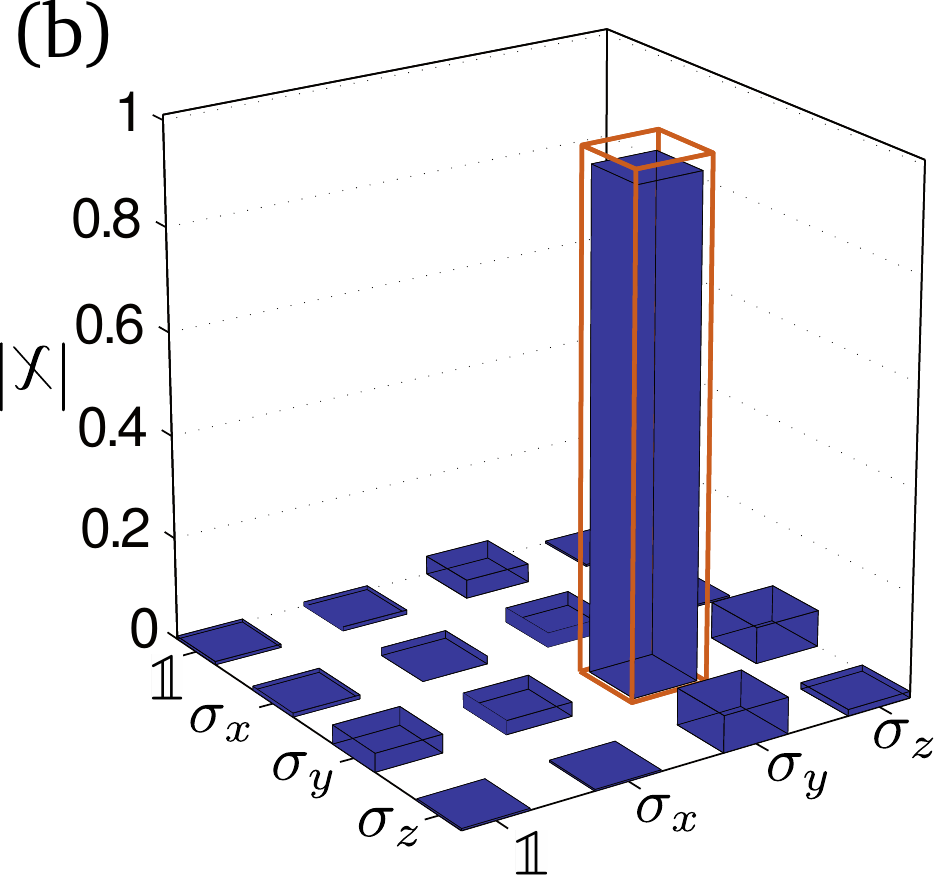}
\includegraphics[height=2.8cm,width=2.8cm]{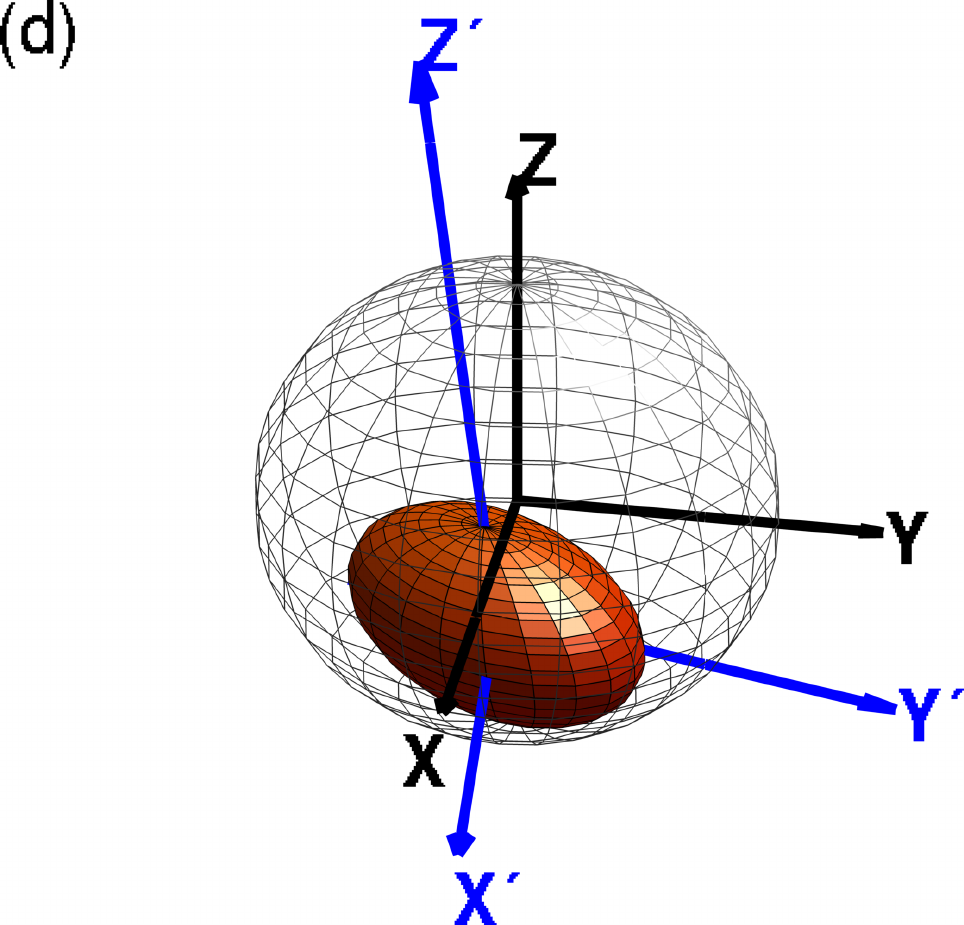}
\includegraphics[height=2.8cm,width=2.8cm]{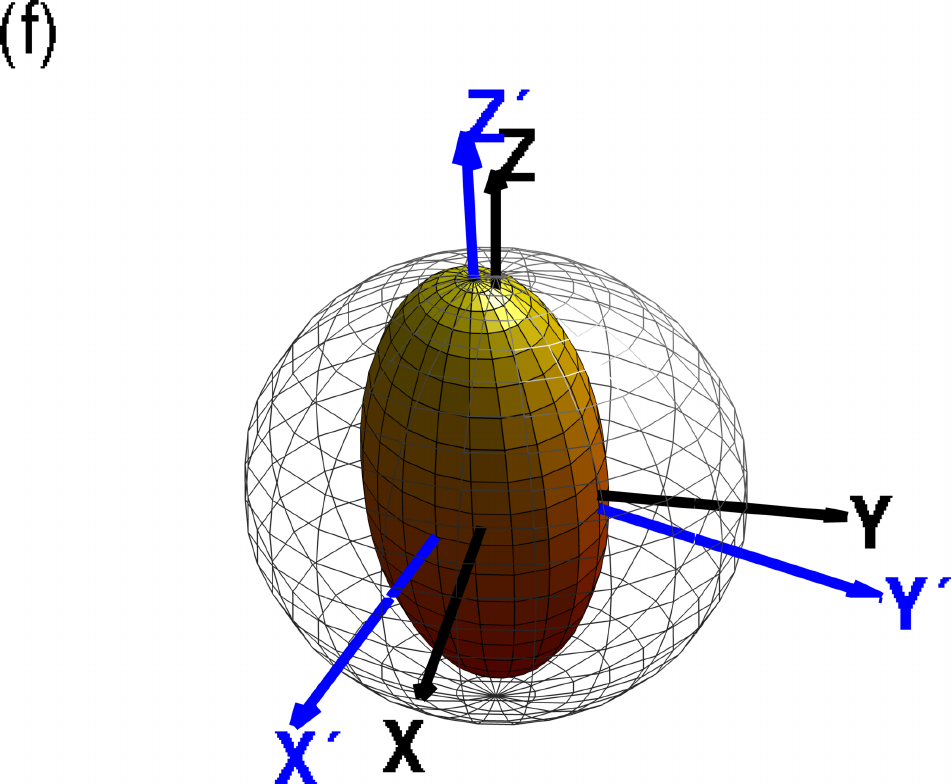}
\fi
\caption{(color online) Experimental results of DCQD for unitary and decoherence
    processes. (a-b) Results of the measured superoperator
    $\chi$ for the rotation operations $U(\pi,0)$ in
    (a) and $U(\pi,\pi/2)$ in (b). 
    Ideally, the target processes have only nonzero elements at positions indicated by the
    orange-bordered bars.
    (c-d) Bloch sphere representation of the
  ideal (c) and measured (d) amplitude damping process
  with 60\% probability~[22]. (e-f) Bloch sphere representation of
  the ideal (e) and measured (f) phase damping process
  with 60\% probability~[22].  Bloch sphere axes in black evolve into the
    spheroid primed axes in blue. A slight imperfection due to residual
    light on the ancilla ion can be observed as a rotation of the spheroids in
    the measured decohering processes.}
\label{figure_results}
\end{figure}

We demonstrate the DCQD method by characterizing the full quantum
process of implemented unitary rotations $\sigma_{x}$ and $\sigma_{y}$
as well as non-unitary processes such as amplitude- and
phase-damping~\footnote[\suppl{}]{See appendix.}. The
$\chi$ matrices reconstructed from the measured probabilities, are
shown in Fig.~\ref{figure_results}(a,b) for $\sigma_x$ and $\sigma_y$
rotations. A single-qubit process can also be visualized by
transforming the pure states lying on a Bloch sphere. In this
Bloch sphere representation, decohering
processes map the unit Bloch sphere (shown as a transparent mesh) to
an ellipsoid of smaller volume~\cite{Chuang}. Implemented amplitude-
and phase-damping processes taking place with a 60\%
probability are shown in this representation in Fig.~\ref{figure_results}(d,f)~\footnote[\suppl{}]{See appendix.}. 
For each input state the experiment was repeated up to 250 times for statistical averaging. 
All processes were reconstructed with a maximum likelihood algorithm to
ensure trace preservation and positivity of the superoperator
$\chi$~\cite{Jezek}. The fidelity $F$ of a process describes the overlap between the measured
$\chi_{meas}$ and the ideal superoperator $\chi_{id}$. For each
process we calculate the overlap between $\chi_{meas}$ and $\chi_{id}$
using the Uhlmann-Jozsa fidelity with the Jamiolkowski
isomorphism~\cite{Jezek, Uhlmann-Jozsa}. Table~\ref{table_fidelity} shows the
calculated fidelities for the implemented DCQD and for SQPT.
The uncertainty in the fidelity was estimated by
parametric bootstrapping based on projection noise in our
measurement~\cite{bootstrapping}.

\begin{table}[h!]
\begin{center}
\begin{tabular}{c @{\hspace{5mm}} c @{\hspace{5mm}}c}
\hline
\hline
Target process & DCQD, F (\%) &
  SQPT, F (\%)\\
\hline
$\mathbb{1}$ & $97.5\pm0.6$ & $98.1\pm1.3$  \\
$\sigma_{x}$ & $96.5\pm1.0$ & $98.1\pm1.3$  \\
$\sigma_{y}$ &$96.6\pm1.4$ & $97.5\pm1.4$  \\
amplitude damping & $95.3\pm1.9$ & $95.2\pm2.7$  \\
phase damping & $97.4\pm0.8$ & $95.7\pm0.8$  \\
\hline
\hline
\end{tabular}
\caption{Calculated process fidelities
$F$ between implemented and target processes as characterized with DCQD and SQPT. All processes were measured with a total of 1000
experimental cycles, which correspond to 1000/4 cycles per
experimental configuration for DCQD and 1000/(4$\times$3)$\sim$84 for
SQPT. The SQPT of the phase damping process was measured with a total of 3000 experimental cycles.}
\label{table_fidelity}
\end{center}
\end{table}

Full QPT of a single-qubit process is also possible with a single
experimental configuration by using additional ancillas and a
generalized measurement (GM). Here, we expand the dimension of the
Hilbert space $H_{A}\otimes H_{S}$, with the system Hilbert space
$H_{S}$ and the ancilla Hilbert space $H_{A}$, such that the dimension
of the total Hilbert space is equal to the number of free
parameters in the process matrix $\chi$~\cite{Mosheni_Resource}. For a
single-qubit process one has to determine all 16 superoperator
elements $\chi_{m,n}$ which leads to an 8 dimensional ancilla Hilbert
space. Therefore we used three ancilla qubits $A_{1}$, $A_{2}$ and $A_{3}$
to quantify a full process $E$ acting on the system qubit S. This GM
is realized by entangling the system and ancilla qubits using
many-body interactions~\cite{ms, ms_exp}, then applying the process
$E$ on S and finally performing BSMs on two pairs.
Figure~\ref{process_scheme}(b) shows the sequence implemented for this
GM which proceeds as follows. First, we create an entangled input
state using maximally and non-maximally entangling M$\o$lmer -
S$\o$renson interactions in combination with global and addressed
single-qubit rotations. After applying the process $E$ on S we perform
a pairwise BSM on the combined output state by implementing two
non-maximally entangling operations $MS(\frac{\pi}{4})$ and two
addressed AC-Stark pulses $U_{Z}^{(1)}(\pi)$ and $U_{Z}^{(3)}(\pi)$,
which separate the entangled system $H(S,A_{1},A_{2},A_{3})$ into a
product state of two subsystems $H(A_{1},A_{3})\otimes
H(S,A_{2})$. These operations are equivalent to two pairwise maximally entangling
gates $MS(\frac{\pi}{2})$ acting on the two subsystems $H(A_{1},A_{3})$ and
$H(S,A_{2})$. The 16 results of the measurement are directly linked to
the 16 superoperator elements $\chi_{m,n}$ by a matrix $\Lambda$
similar to Eq.~(2). Using this technique we reconstructed unitary
processes $\{\mathbb{1}, \sigma_{x}=U(\pi,0), \sigma_{y}=U(\pi,\frac{\pi}{2}), \sigma_{z}=U_{z}^{(1)}(\pi)\}$ acting
on a single qubit with a fidelity of $\{99.70\pm 0.02, 97.30\pm 0.29,
99.80\pm 0.01, 99.40\pm 0.02\}\%$. All processes were measured with a total of 5000 cycles.

\begin{figure*}[t!]
\iffigs\includegraphics{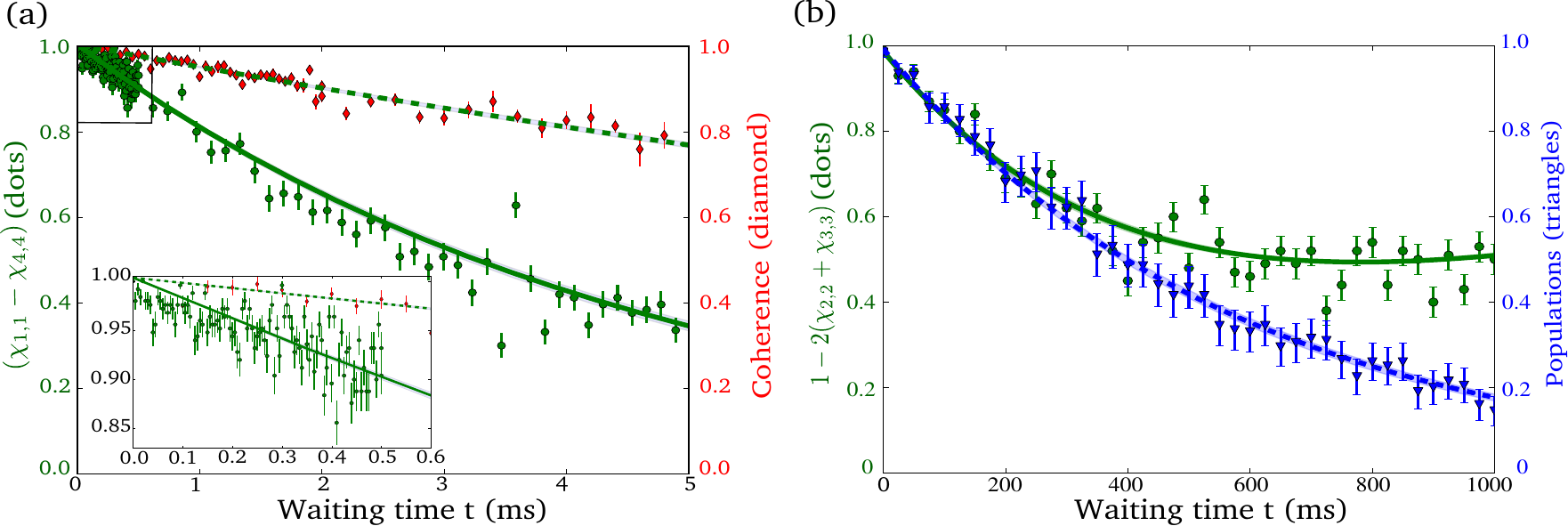}\fi
\caption{(color online) Simultaneous measurement of phase decoherence (a) and the
spontaneous decay (b) of a two-qubit system. The \method{} technique
(green dots) is compared to a Ramsey-contrast measurement (red
diamonds) and a spontaneous-decay measurement (blue triangles) (see text). 
The measurement using the \method{} method in (a) was carried
out on the entangled two-qubit system
(\mbox{$\exp(-\frac{4t}{T_{2}^{DCRT}})$} scaling) whereas the red diamonds
were measured on a single qubit with the Ramsey-contrast technique
(\mbox{$\exp(-\frac{t}{T_{2}^{trad}})$} scaling). The shaded areas correspond to the envelope of the
curves with the decay times $T_{1,2}^{DCRT,trad}\pm \Delta
T^{DCRT,trad}_{1,2}$, considering the statistical errors $\Delta
T_{1,2}^{DCRT,trad}$. The relaxation time measurements, using the
\method{} method and, in comparison, the traditional Ramsey-contrast
and spontaneous decay measurement, yield:
T$_2^{DCRT}$=18.8(5)~ms, T$_2^{trad}$=19.4(8)~ms,
$\mathrm{T}_{1}^{\method{}} = 1130(47)$~ms and $\mathrm{T}_{1}^{trad}
= 1160(30)$~ms.} \label{decay}
\end{figure*}

In contrast to previous QPT measurements of engineered processes, the
process of phase (amplitude) damping occurs naturally in our
system due to magnetic field fluctuations (spontaneous
decay)~\cite{monz}. The dynamical parameters T$_1$ and T$_2$ can,
however, be determined simultaneously with only the first input state
$\rho_1$ being subject to the DCQD scheme even if the damping
processes act collectively on both qubits (as in our experimental
system~\cite{monz}). This method, named \method{} above, consists of
preparing an input Bell state
$\rho_1=|\Phi^{+}\rangle\langle\Phi^{+}|$, exposing both qubits to the
damping processes for a time $t$ and a final BSM, which yields the
diagonal elements $\chi_{i,i}$ of the process matrix. As described in
the supplementary material~\footnote[\suppl{}]{See appendix.} and assuming Markovian noise the
dynamical parameters are then given by

\begin{eqnarray}
e^{-\frac{N^{2}\mathrm{t}}{\mathrm{T}_{2}}} &=& \chi_{1,1} - \chi_{4,4} \\
&=&\mathrm{Tr}\left\{\left[|\Phi^{+}\rangle\langle\Phi^{+}| - 
|\Phi^{-}\rangle\langle\Phi^{-}|\right]\mathrm{E}\left(|\Phi^{+}\rangle\langle\Phi^{+}|\right)\right\}, \nonumber
\end{eqnarray}
\begin{multline}
1+2e^{-\frac{2\mathrm{t}}{\mathrm{T}_{1}}}-2e^{-\frac{\mathrm{t}}{\mathrm{T}_{1}}} = 1-2(\chi_{2,2} + \chi_{3,3}), 
\end{multline}

\noindent with $N$ the number of ions. From the entries of the $\chi$ matrix
corresponding to $\mathbb{1}$ and $\sigma_{z}$ ($\sigma_x$ and
$\sigma_y$) operations, T$_2$ (T$_1$) depends on the probability that
no error or phase flips (bit flips) occur on the entire system. A fit
of \method{} measurements $\chi_{i,i}$ to Eqns.~(3-4) at different
times $t$ thus yields T$_1$ and T$_2$ using a single experimental
configuration. We explored this \method{} technique in our
experimental system. The measurement results of the decoherence
estimation are shown in Fig.~\ref{decay}(a). The green dots show the
difference between the diagonal elements $\chi_{1,1}$ and $\chi_{4,4}$
as a function of the waiting time $t$. The spontaneous decay
of the system is shown in Fig.~\ref{decay}(b) by plotting
$1-2(\chi_{2,2}+\chi_{3,3})$ as a function of time. For every waiting
time $t$ the experiment was repeated up to 250 times to gain significant
statistics.

We can compare the \method{} technique with two traditional methods
that use product input states: Ramsey-contrast measurements for
phase-decoherence estimation and direct spontaneous-decay
measurements~\cite{Ramsey}. A Ramsey-contrast measurement is realized
by initializing the ion in the state $(|0\rangle+|1\rangle)/\sqrt{2}$
by a global rotation $U(\frac{\pi}{2},0)$, followed by a waiting time
$t$ and finally applying a second rotation $U(\frac{\pi}{2},\phi)$ in
which the phase $\phi$ is varied. The observed contrast as a function
of $\phi$ corresponds to the preserved
phase coherence. Spontaneous-decay measurements, instead, consist of
measuring the probability of detecting both ions in the excited state
$|0\rangle$ as a function of time. The results of these
Ramsey-contrast (spontaneous-decay) measurements are shown in
Fig.~\ref{decay}(a) (Fig.~\ref{decay}(b)) as red diamonds (blue
triangles). The measured relaxation times corresponding to the
traditional methods are called $\mathrm{T}_{1}^{trad}$ and
$\mathrm{T}_{2}^{trad}$. The exponential fit (green line)
of Eq.~(3) to the data was estimated with $N=2$ (collective
dephasing) and yields T$_2^{DCRT}$=18.8(5)~ms. The Ramsey-contrast measurements (red diamonds)
were carried out on a single ion and yield a coherence time of
T$_2^{trad}$=19.4(8)~ms. The green dotted line in Fig.~\ref{decay}(a)
corresponds to the fitted function to Eq.~(3) (green line) with $N=1$
instead of $N=2$ and shows good agreement with the single-ion
Ramsey-contrast measurement. Therefore the \method{} technique enables
the characterization of the phase decoherence of the collective system
(green line) and also gives a conclusion about the phase decoherence
of a single ion (green dotted line).  An exponential fit of the decay
data of Fig.~\ref{decay}(b) to Eq.~(4) gives the characteristic
lifetime $\mathrm{T}_{1}^{\method{}} = 1130(47)$~ms for the \method{}
technique (green line) and $\mathrm{T}_{1}^{trad} = 1160(30)$~ms for
the traditional method (blue dotted line), which are in good agreement
with previously measured values~\cite{Staanum} of 1148(18)~ms.\\ 

In summary, we have experimentally demonstrated two different
approaches for the full characterization of single-qubit quantum processes,
lowering the required experimental configurations from 12 to 4 using DCQD
and a single configuration via the GM method. The reconstruction of coherent
and incoherent processes was shown with fidelities of $\ge$~97\% using
DCQD. In particular, we have observed a lower statistical
uncertainty of the fidelity of some of the processes compared to the SQPT.
Nevertheless, a matter of further investigations is a comparison
of the scaling in the number of experimental cycles required for the SQPT
and DCQD to achieve a target uncertainty in the fidelity (e.g. see identity 
process in Table~\ref{table_fidelity}). 
The \method{} technique, based on the DCQD protocol, was used as a
powerful tool to characterize the noise in our system by measuring the
relaxation times T$_{1}$ and T$_{2}$ simultaneously with one
experimental setting. This technique indicates good agreement with traditional
methods as Ramsey-contrast and spontaneous decay measurement. In principle, there is an
improvement of a factor of two in the measurement time if T$_{1}$ is
of the same order of magnitude as T$_{2}$, which is not the case for
our setup. In contrast, spin-based solid state systems are
collectively affected by noise and T$_{1}$$\approx$T$_{2}$, which
would lead to a significant improvement of the measurement
time~\cite{dots_decoherence}. Another application of \method{} could
be for biological systems where dissipative dynamics play a crucial
role~\cite{quantum_transport_Masoud, QPT_chemical}. The same measurement procedure can also
be used as a tool to quantify Hamiltonian parameters
efficiently~\cite{Mohseni_parameters, Mosheni_DCQD_2009}.

\begin{acknowledgments}
We gratefully acknowledge support by the Austrian Science Fund (FWF), through the Foundations and 
Applications of Quantum Science (SFB FoQus), by the 
European Commission AQUTE, by IARPA, QuISM MURI and DARPA QuBE Program as well as the Institut f\"ur Quantenoptik und Quanteninformation GmbH. 
Julio T. Barreiro acknowledges support by a Marie Curie International 
Incoming Fellowship within the 7\textit{th} European Community Framework
Programme.   
\end{acknowledgments}

\pagebreak
\appendix

\section{Appendix}

\section{Implementation of amplitude- and phase-damping}

\begin{center}
\begin{figure*}[t!]
\iffigs\includegraphics[scale=2]{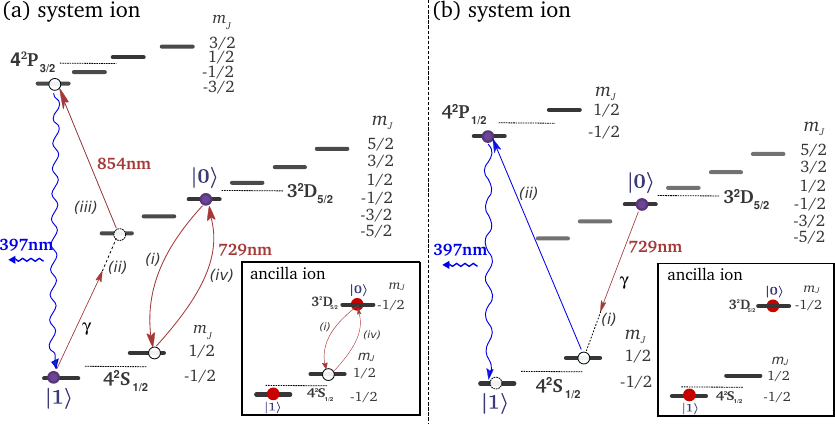}\if
\caption{\label{damping_scheme}(color online) Schematic of the phase
  damping (a) and amplitude damping (b) processes acting on the system ion S: (a) Transfer of the $S_{1/2}(m
  = -1/2)$ population with a probability of $\sin(\gamma)^{2}$ to
  the $P_{3/2}$ state followed by optical repumping to the ground state while
  the population of the $D_{5/2}(m = -1/2)$ state is hidden in the $S_{1/2}(m =
  1/2)$ state. (b) The amplitude damping process is carried out by coherently transfering
  the population of the $D_{5/2}(m = -1/2)$ state to the $S_{1/2}(m = 1/2)$
  state of the system ion followed by optical pumping to the ground state $S_{1/2}(m = -1/2)$. The
  ancilla ion is not affected during the whole processes, as shown in the inset figures.}
\end{figure*}
\end{center}

For the complete quantum process tomography using DCQD it was considered  
that the process of interest only acts on the system qubit.
In general due to residual light on the ancilla ion (crosstalk), it is experimentally challenging to apply single-qubit operations
on one ion without affecting the second one.  
The unitary operations on the system ion S without affecting the ancilla ion A is experimentally implemented 
by the following refocusing technique: (i) Applying a global rotation $U(\frac{\pi}{2},\phi)$ around the $X$ or $Y$
axis of the Bloch sphere. (ii) Shifting the phase of the second ion by $\pi$
via an AC-Stark pulse $U_{Z}^{(2)}(\pi)$. (iii) Repetition of the first step. 
This effectively reduces the crosstalk, for single qubit operations roughly 
by a factor of 6, ideally to less than 1\%.\\
The process of \textbf{phase damping} is implemented by the following steps (see Fig.~\ref{damping_scheme}(a)): (i) Hiding the population
of the $D_{5/2}(m = -1/2)$ state of both ions by applying a $\pi$-pulse on the
$D_{5/2}(m = -1/2) \leftrightarrow S_{1/2}(m = 1/2)$ transition. (ii) Transfering a
certain amount of the population of the $S_{1/2}(m = -1/2)$ state to the
$D_{5/2}(m = -5/2)$ state. The excitation probability is experimentally
controlled by the pulse length $t$ and is given by $\sin(\gamma)^{2}$, with
$\gamma = \frac{\pi t}{T}$ and $T$ the Rabi-oscillation period time. (iii)
Repumping the transfered population into the $P_{3/2}$ state by a laser pulse at
a wavelength of 854 nm whereupon the system qubit decays spontaneously into the
ground state. (iv) Finally the second step is repeated to reverse the hiding
process. Because of the spontaneous decay after the third step the state loses
its phase information. The hiding process described in the first step is
necessary to prevent the $D_{5/2}(m = -1/2)$ state from not being affected by the
repumping process which would have the same effect as amplitude damping.\\
\textbf{Amplitude damping} is carried out within two steps (see Fig.~\ref{damping_scheme}(b)). First a certain
amount of population is coherently transfered from the $D_{5/2}(m = -1/2)$ to
the $S_{1/2}(m = 1/2)$ state by an addressed laser pulse analog to step (ii) of
the phase damping process. Then the whole population of the $S_{1/2}(m = 1/2)$
state is transfered to the $S_{1/2}(m = -1/2)$ ground state by an optical
pumping process performed by a $\sigma^{-}$-polarized laser beam at 397~nm
exciting the ion first to the $P_{1/2}(m = -1/2)$ state followed by spontaneous 
decay to the $S_{1/2}(m = -1/2)$. The ancilla ion is not affected by the optical 
pumping process because the $\sigma^{-}$-polarized pulse can only excite the $S_{1/2}(m = 1/2)$ ground
state. 

\section{Direct characterization of collective longitudinal and transverse relaxation processes}

Let us consider a global quantum \textit{homogenization or
thermalization} process acting on some qubits for a time $t.$ Thus
for each single-qubit density matrix $\rho$, with $\rho_{00}=a$ and
$\rho_{01}=b$ in the
computational basis, evolves into the state $\rho(t)$ with $%
\rho_{00}(t)=(a-a_{0})\exp(-t/T_{1})+a_{0}$ and $\rho_{01}(t)=b%
\exp(-t/T_{2}) $, where $T_{1}$ and $T_{2}$ ($T_{2}\leqslant2T_{1}$)
are relaxations and dephasing time-scales of the system,
respectively. That is, the system approaches an equilibrium state
identified by $a_{0}\in [0,1]$.

\noindent In order to measure these time scales using the DCQD approach
we need to create quantum correlations between
each pair of neighboring qubits, e.g., in the form of a Bell-state 
$\rho=|\Phi^{+}\rangle\left\langle \Phi^{+}\right\vert$ where 
$|\Phi^{+}\rangle =(|00\rangle+|11\rangle)/\sqrt{2},$ and then let
all of the qubits evolve for a time t leading to state
$\mathcal{E}(\rho).$ Thus by performing a Bell-state measurement
between the same neighboring qubits we obtain (\cite{Mohseni_parameters},~\footnote{M. Mohseni, in preparation, 2012}):

\begin{equation*}
Tr[P^{11}\mathcal{E}(\rho)]-Tr[P^{44}\mathcal{E}(\rho)]=e^{-\frac {2t
}{T_{2}}}
\end{equation*}

\noindent where $P^{kk^{\prime}}=\left\vert B^{k} \right\rangle \left\langle
B^{k^{\prime}}\right\vert$, and $|B^{k}\rangle$ for $k=1,2,3,4$
corresponds to the Bell-states $\left\vert \Phi^{+}\right\rangle $,
$\left\vert \Psi ^{+}\right\rangle $, $\left\vert
\Psi^{-}\right\rangle $, and $\left\vert \Phi^{-}\right\rangle $,
respectively, where $|\Phi^{\pm}\rangle
=(|00\rangle\pm|11\rangle)/\sqrt{2},~|\Psi^{\pm}\rangle=(|01\rangle
\pm|10\rangle)/\sqrt{2}$. Using the fact that $Tr[P^{11}\mathcal{E}^{T}(\rho)]=\chi_{1,1}$ and
$Tr[P^{44}\mathcal{E}^{T}(\rho)]=\chi_{4,4}$ we obtain:

\begin{eqnarray}
e^{-\frac{\mathrm{2t}}{\mathrm{T}_{2}}} &=& \chi_{1,1} - \chi_{4,4}.
\end{eqnarray}

\noindent It is important to mention that this expression is only valid for
Markovian noise acting on the entire system. For our system it
was observed \cite{monz} that the phase decoherence of Greenberg-Horne-Zeilinger (GHZ) 
states scales with $\exp(-N^{2}t)$ instead of $\exp(-Nt)$, with N the number of ions. Therefore Eq.~(1) has to be modified, which
leads to the following result for the phase-decoherence:

\begin{eqnarray}
e^{-\frac{\textrm{N}^{2}\mathrm{t}}{\mathrm{T}_{2}}} &=& \chi_{1,1} - \chi_{4,4}.
\end{eqnarray}
    
\noindent In the case of T$_{1}$, the outcome of a BSM within the DCQD formalism yields:

\begin{eqnarray}
1-2(Tr[P^{22}\mathcal{E}(\rho)]+Tr[P^{33}\mathcal{E}(\rho)])=\nonumber \\ 
= (1-2a_{0})^{2}(1-2e^{-\frac{t}{T_{1}}})+(2+4a_{0}(a_{0}-1))e^{-\frac{2t}{T_{1}}}.
\end{eqnarray}

\noindent For our system the equilibrium state is described by the ground state and 
therefore $a_{0}=1$, which leads to: 

\begin{eqnarray*} 
1-2(\chi_{22}+\chi_{33})= (1-2e^{-\frac{t}{T_{1}}})+2e^{-\frac{2t}{T_{1}}}.
\end{eqnarray*}

\noindent For a unital quantum \textit{homogenization} process 
($\mathcal{E}(I)=I$), we have $a_{0}=\frac{1}{2}$ (i.e., a completely stochastic
equilibrium state). Thus the relation for T$_{1}$ becomes:

\begin{equation*}
e^{-\frac{2\mathrm{t}}{\mathrm{T}_{1}}} = 1-2(\chi_{2,2} + \chi_{3,3}).
\end{equation*}\\
 
It is remarkable that even if we use the relations developed for the
ideal DCQD scheme, assuming T$_{1}$ and T$_{2}$ acting only on the
system qubit (considering the other system as a reference signal or
a noiseless ancilla), we can still obtain both T$_{2}$ and T$_{1}$
only to be smaller than the actual value by a factor of 2.
Note that, due to orthogonality of the BSM outcomes, it is easy to
unambiguously distinguish T$_{1}$ from T$_{2}$. Traditionally, in
order to measure the longitudinal and transverse relaxation times,
one needs to measure two non-commutative observables 
(e.g., Pauli operators $\sigma_{z}$ and $\sigma_{x}$) on two sub-ensembles of identical systems.

\end{document}